\documentclass{aa} 

\usepackage{graphicx}
\usepackage{natbib}         
\usepackage{txfonts}
\usepackage{xcolor}
\usepackage{hyperref}
\usepackage{dblfloatfix}

\begin{document} 
   \title{Invariant manifolds in barred galaxy simulations}

    \subtitle{I. Material density waves}

   \author{T. Soler-Terricabras
          \inst{1}\fnmsep\inst{2}\fnmsep\inst{3},
          M. Romero-Gómez\inst{1}\fnmsep\inst{2}\fnmsep\inst{3}
          \and
          S. Roca-Fàbrega\inst{4}
          }
    \authorrunning{Soler-Terricabras et al.}

   \institute{Departament de Física Quàntica i Astrofísica (FQA), Universitat de Barcelona (UB), c. Martí i Franquès, 1, 08028 Barcelona, Spain\\
              \email{tsoler@fqa.ub.edu}
        \and
             Institut de Ciències del Cosmos (ICCUB), Universitat de Barcelona (UB), c. Martí i Franquès, 1, 08028 Barcelona, Spain
        \and
             Institut d’Estudis Espacials de Catalunya (IEEC), Edifici RDIT, Campus UPC, 08860 Castelldefels (Barcelona), Spain
        \and 
             Lund Observatory, Division of Astrophysics, Department of Physics, Lund University, SE-221 00 Lund, Sweden
             }

   \date{Received XXX; accepted YYY}

  \abstract
   {The origin and nature of spiral arms in barred galaxies are still under debate. Among others, the theory of invariant manifolds provides a framework capable of explaining spiral and ringed structures in these systems.}
   {We investigate the dynamical origin and kinematic signatures of spiral structure in an $N$-body simulation of an isolated barred galaxy, assessing whether invariant manifold theory provides a consistent dynamical framework to disentangle the disc particle populations and to identify those that genuinely build, trace, and sustain the spiral arms.}
   {For each simulation snapshot, we measure the bar and spiral properties and reconstruct the corresponding gravitational potential using \texttt{AGAMA}. We compute the Jacobi energy of disc particles and classify them relative to the energies of the equilibrium points, thereby isolating manifold-compatible orbits. We analyse their spatial distribution and velocity structure to characterise spiral-related streaming motions.}
   {The Jacobi constant provides a physically motivated dynamical separator that reveals three distinct kinematic populations: (i) low-energy particles on nearly circular orbits populating most of the disc ($\sim60\%$), (ii) high-energy particles associated with banana orbits ($\lesssim10\%$), and (iii) manifold-compatible particles ($\sim30$--$40\%$) originating near the bar and following transit orbits along the spiral arms. Only the manifold-compatible population generates the prominent outward-migrating ridge observed in the $R-v_\varphi$ plane and reproduces the characteristic spiral streaming pattern. In contrast, the low-energy population exhibits a global quasi-circular motion with small perturbations induced by the self-gravity of the spiral structure.}
   {Our results demonstrate that the spiral arms are dynamically traced by the manifold-compatible population, which forms the backbone of the structure and drives effective radial transport. The bulk of low-energy disc particles responds to the spiral perturbation similarly to the traditional density wave picture, enhancing the density contrast caused by the invariant-manifold compatible particles. In this framework, barred spiral arms emerge as material structures sustained by manifold-guided transport, with the surrounding disc behaving as a system of material density waves.}


    \titlerunning{Invariant manifolds in barred galaxy simulations -- I}
    \authorrunning{T. Soler-Terricabras et al.}

   \maketitle

\section{Introduction}\label{sec:introduction}
The physical origin of spiral arms in barred galaxies remains a matter of on-going debate \citep[e.g.][]{Sellwood2011, DobbsBaba2014, Baba2015, Grand2026}.
In particular, the dynamical interplay between bars and spiral arms is still not fully understood.
While bars are generally long-lived, self-consistent non-axisymmetric structures supported by well-defined orbital families, it is still unclear whether the spiral arms that emerge from them are similarly long-lived density waves or instead recurrent, transient patterns.
Distinguishing between these scenarios requires more than reproducing the observed morphology of spiral structure; it demands a dynamical understanding of the orbital mechanisms that support the arms.
Spiral arms are not merely surface-density enhancements; both observations and simulations show that they are kinematically perturbed regions of the disc, characterised by systematic streaming motions and coherent velocity patterns that set them apart from the surrounding stellar population \citep[e.g.][]{RocaFabrega2014, Eilers2020}. 
Any successful dynamical framework for spiral structure must therefore account simultaneously for their morphological prominence and their kinematic signatures.

Classical density wave theory {for non-barred spiral galaxies} predicts predicts specific kinematic imprints associated with spiral structure \citep[e.g.][]{LinShu1964, LinShu1966, LinShu1969, Yuan1969, Kalnajs1973}. 
However, in barred galaxies, the theory of invariant manifolds \citep{MRG2006, MRG2007, Voglis2006a, Voglis2006b, Athanassoula2009a, Athanassoula2009b, Athanassoula2010} provides a particularly compelling phase-space interpretation. 
In this framework, spiral arms are not {prescribed beforehand} but emerge {self-consistently} as the configuration-space manifestation of invariant manifolds associated with the unstable Lyapunov orbits about Lagrangian points of the bar. 
The orbits that support the spiral structure are therefore guided by these manifolds, establishing a direct dynamical link between the bar and the spiral arms.

A major advance in constraining such theoretical frameworks has come from the high-precision astrometric data delivered by the \textit{Gaia} mission \citep[ESA,][]{Gaia2016}.
One of its most remarkable findings has been the discovery of rich phase-space substructure in the Galactic disc {of the Milky Way}. 
In particular, the distribution of stars in the Galactocentric radius–azimuthal velocity plane ($R$–$v_\varphi$) exhibits prominent diagonal overdensities, commonly referred to as \textit{ridges}, first characterised in Gaia DR2 by \citet{Antoja2018}. 
Subsequent studies have shown that these ridges are robust features of the Milky Way disc and likely reflect underlying dynamical mechanisms rather than observational artefacts \citep[e.g.][]{Minchev2009, Ramos2018, Martinez-Medina+2019, Fragkoudi2019, Laporte2019, GaiaCollaboration2021, Antoja2022}.
Various interpretations have been proposed for the origin of ridges, including phase mixing induced by transient spiral arms \citep{Kawata2018, Hunt2019, Khanna2019}, resonant effects of the bar \citep{Monari2019}, and combined bar–spiral interactions \citep{Hunt2018b, Hunt2019}.
{Interestingly}, the correlation between ridges in $R$–$v_\varphi$ and arches in $v_R$–$v_\varphi$ reveals that both are projections of the same six-dimensional phase-space substructure \citep{Bernet2022}. 
This phase-space coherence implies that these features encode information about the orbital architecture of the disc. 
In this context, invariant manifolds --known to organise orbital flows in barred potentials-- naturally emerge as a promising framework to interpret the origin of these correlated structures.

If spiral arms in barred galaxies are indeed governed by invariant manifolds, then their supporting orbits should leave identifiable imprints in phase space. In this context, the ridges observed in the $R$–$v_\varphi$ plane provide a powerful diagnostic: they may constitute the kinematic signature of stars guided by manifold structures.
In this paper, we investigate this possibility by analysing the kinematic imprints of spiral arms in the disc of an $N$-body simulation, and by examining whether the ridges in the $R$–$v_\varphi$ plane can be understood within the framework of invariant manifold theory. 
Specifically, we aim to determine whether the stars populating these ridges are dynamically connected to the manifold-driven spiral arms, thereby establishing a direct link between configuration-space morphology and phase-space structure.

This paper is structured as follows. 
We begin by outlining the theoretical basis from invariant manifolds frameworks and the fundamental equations relevant to our analysis in Sec.~\ref{sec:theory}. 
We then describe the $N$-body simulation examined in this study in Sec.~\ref{sec:methodology}, where we also introduce the methodology developed to construct snapshot-specific gravitational potentials and to evaluate particle dynamics in the rotating frame. This approach enables us to characterise the key dynamical parameters of each snapshot and to obtain a consistent model of the gravitational potential at every stage of the simulation.
In Sec.~\ref{sec:kinematic}, we identify the kinematic signatures associated with the spiral structure in this barred galaxy model and assess their consistency with the predictions of manifold theory. We then define, in Sec.~\ref{sec:populations}, a selection criterion based on the Jacobi constant, $E_J$, to separate dynamically distinct orbital populations. The dynamical properties of these groups are analysed and compared with those of the spiral arms.
{In Sec.~\ref{sec:discussion}, we discuss the origin and robustness of kinematic ridges, with particular emphasis on the respective roles of the bar and spiral structure. 
In Sec.~\ref{sec:caveats}, we comment on the main caveats and limitations of our approach.} 
Finally, we summarise the main conclusions of this work in Sec.~\ref{sec:conclusions}.
 
\section{Theoretical basis}\label{sec:theory}
    Within the context of barred galaxies, it is often convenient to work in the non-inertial reference frame that rotates with the bar at pattern speed $\mathbf{\Omega_b}=\Omega_\text{b}\mathbf{e_z}$, with $\Omega_\text{b}>0$ by convention --which corresponds to counter-clockwise rotation-- due to the rigid-body nature of the bar.
    Let $(R,\varphi)$ denote inertial polar coordinates and $(r,\phi)$ the coordinates in the frame corotating with the bar.
    Then, $r=R$ and $\phi=\varphi-\Omega_\text{b}t$.
    In the rotating frame, the planar motion of particles is governed by the Hamiltonian
    \begin{equation}\label{eq:jacobi}
        E_J\equiv\mathcal{H}(r,\phi,v_r,v_\phi)=\frac{1}{2}(v_r^2+v_\phi^2)+\Phi_\text{eff}(r,\phi)
    \end{equation}
    where $v_r$ and $v_\phi$ are radial and tangential velocities, and the effective gravitational potential, $\Phi_\text{eff}$, is defined as
    \begin{equation}\label{eq:poteff}
        \Phi_\text{eff}(r,\phi) \equiv \Phi(r,\phi) - \frac{1}{2}\Omega_\text{b}^2\,r^2
    \end{equation}
   The structure of $\Phi_\text{eff}$ in a barred galaxy typically features five equilibrium points in the rotating frame, also called \textit{Lagrangian points}, denoted by $L_i$, for $i\in\{1,...,5\}$, which are defined as the loci where $\mathbf{\nabla}\Phi_\text{eff}(x,y)=0$.  
   The $\Phi_\text{eff}$ minimum, namely the point $L_3$, lies at the origin and is surrounded by the $x_1$ family of periodic orbits \citep[e.g.][]{Contopoulos1980}, which constitute the backbone of the bar \citep[e.g.][]{Athanassoula1984}.
   $L_4$ and $L_5$, located symmetrically along the bar’s semi-minor axis {in symmetric bar potentials}, are $\Phi_\text{eff}$ maxima and each one is surrounded by families of periodic orbits known as \textit{banana orbits} due to their characteristic shape \citep{Contopoulos1978, Contopoulos1980, Contopoulos2002}.
   The remaining two equilibrium points, $L_1$ and $L_2$, are unstable saddle points located near the edges of the bar along its semi-major axis, and are also symmetric with respect to the origin \citep{BinneyTremaine}.
   Around each of them there exists a family of periodic orbits referred to as Lyapunov orbits, or $\ell_1$ \citep{Lyapunov1949}.
   These are elongated, almost elliptical orbits whose major axes are {parallel to} the bar’s semi-minor axis. 
   They appear for Jacobi constant values larger than those associated with the corresponding equilibrium point, $E_{L_i}$, $i\in\{1,2\}$, and grow in size while remaining nearly nested as energy increases \citep{Skokos2002}.
   Except at much higher energies, these Lyapunov orbits remain unstable, and therefore cannot trap nearby orbits: initial conditions in their vicinity in phase space inevitably diverge from them.
   However, this divergence does not occur arbitrarily. The possible escape directions are determined by the corresponding invariant manifolds, which can be thought of as dynamical tubes guiding the motion of particles with energies compatible to their own's \citep{Danby1965, Gomez2004, MRG2006, Patsis2006, Athanassoula2009a}.
   Theoretically, only particles whose energy is bounded by the Jacobi constants of the saddle-unstable equilibrium points and that of the associated manifolds, can participate in this manifold-driven dynamics \citep{MRG2006,Athanassoula2009a}.

\section{Methodology}\label{sec:methodology}

\begin{figure}[!t]
   \centering
   \includegraphics[width=\hsize]{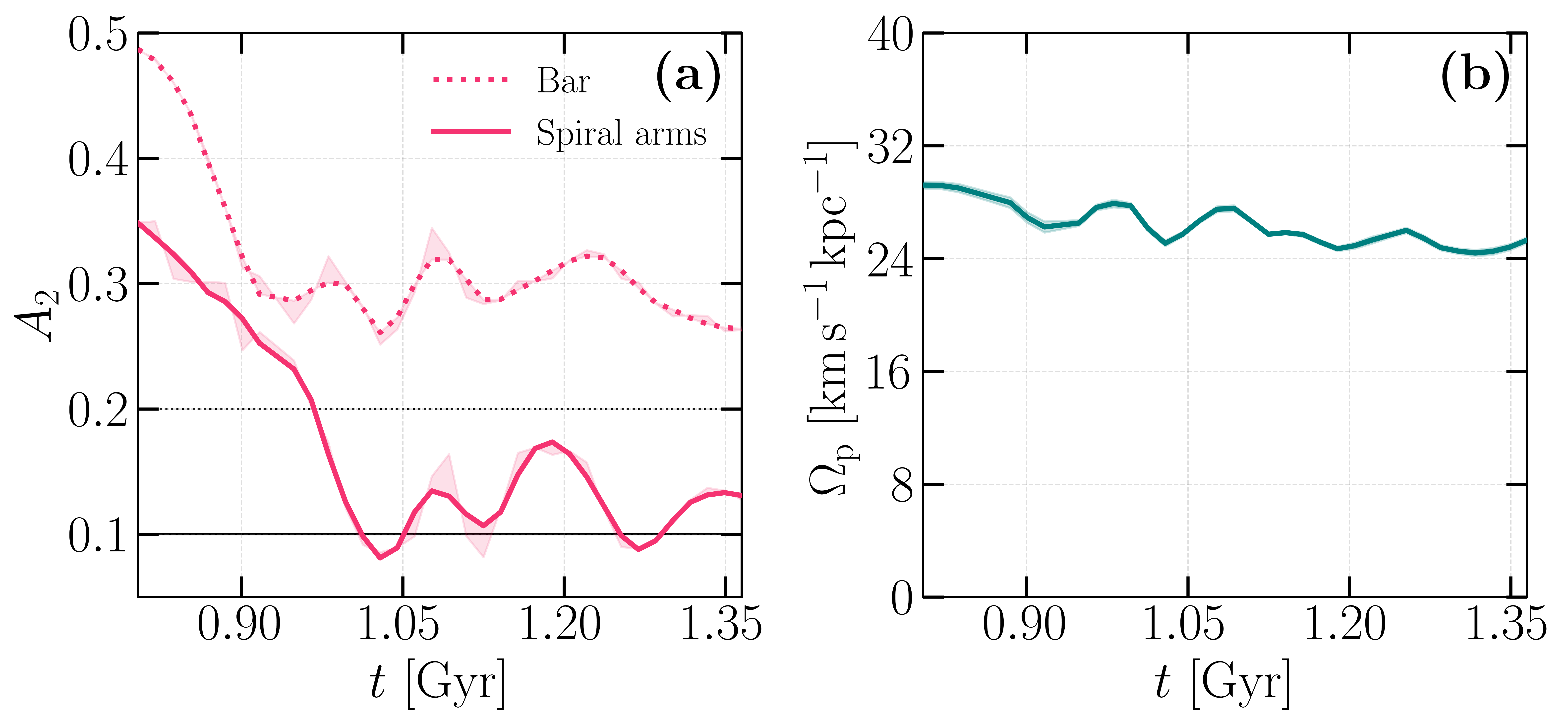}
        \caption{Time evolution of the bar and spiral arms diagnostics for the B1 model, following the method of \citet{Dehnen2022}. 
        Time is measured from the start of the simulation.
        {(a)}: Fourier $m=2$ amplitudes, $A_2$, for the bar (dotted red; averaged over $[R_0,R_1]\,\text{kpc}$, as defined in Appendix B of \citet{Dehnen2022}) and for the spiral arms (solid red; averaged from $R_1$ to 10 kpc).
        The dotted and solid horizontal lines mark $A_2=0.2$ (bar threshold) and $A_2=0.1$ (spiral fading), respectively.
        {(b)}: Bar pattern speed, $\Omega_{\rm p}$, as a function of time.}

         \label{fig:B1_parameters}
   \end{figure}

We analyse the B1 simulation of \citet{RocaFabrega2013}, a pure $N$-body model of an isolated barred galaxy, {i.e. a collisionless simulation including only stellar and dark matter particles, with no gaseous component or star formation processes}. 
The stellar disc has $10^6$ particles accounting for $M_{d}=5.0\times10^{10}\,M_\odot$, and is embedded in a dark matter halo of $M_{h}=1.38\times10^{12}\,M_\odot$ represented by $7.65\times10^6$ particles. 

From the full 4.6\,Gyr run we select 36 snapshots, spanning 561\,Myr during which the bar completes two and a half revolutions, right after disc stabilisation and bar formation.
For this snapshot selection, it was required that the $m=2$ Fourier amplitude in the bar region, namely $[R_0,\,R_1]$ kpc, was $A_2>0.2$; and that a clear bisymmetric spiral structure was present in the disc, i.e. $A_2>0.1$ averaged over $[R_1,10]$\,kpc (see Fig. \ref{fig:B1_parameters}a).
In this context, $R_0$ and $R_1$ denote the inner and
outer edges of the bar region as defined in \citet{Dehnen2022}.
{The choice of an outer limit at $R=10$~kpc for the computation of the $A_2$ amplitude of the arms is motivated by the need to characterise the dominant, large-scale spiral structure while avoiding biases introduced by the low-density outer disc. Beyond this radius, the stellar surface density declines significantly, which can artificially enhance the relative amplitude of Fourier modes and lead to an overrepresentation of weak or fragmented outer features. By restricting the analysis to $R \leq 10$ kpc, we ensure that the measured $m=2$ amplitude reflects the coherent, high-density portion of the spirals that is dynamically connected to the bar. We have verified that our main conclusions do not depend sensitively on this choice, as the inner spiral structure—where the bar–spiral coupling is most relevant—is fully captured within this radial range.}\\

{The analysis presented in this work is therefore based on the analysis of a single $N$-body simulation and a selected temporal window, so its direct applicability is formally restricted to systems sharing similar dynamical properties. 
In this particular set-up, the bar–spiral morphology corresponds to a configuration in which the spiral arms are connected to the bar and share its pattern speed at their connection, as reported in \citet{RocaFabrega2013} (see Fig.~4 there). 
Beyond this region, the spiral pattern speed decreases only moderately with radius further out.
Over the analysed time interval, the bar remains the dominant non-axisymmetric component of the potential and its pattern speed stays approximately constant, whereas the spiral arms exhibit noticeable variations in amplitude (Fig.~\ref{fig:B1_parameters}).}

{Under these conditions, the dynamical framework explored here can be approximated, to first order, by a steadily rotating bar potential. This configuration is not universal in $N$-body simulations, where both pattern speeds and non-axisymmetric components can evolve significantly over time \citep[e.g.][]{Athanassoula2012, Wu2018}.
Accordingly, the present analysis focuses on systems in which the bar dominates the non-axisymmetric structure and evolves on timescales longer than those considered here.}\\

    The first step in this analysis is to determine the pattern speed, $\Omega_\text{b}$, bar angle, $\psi_b$, and bar length, $R_1$, for each snapshot using the algorithm proposed by \citet{Dehnen2022}.
    The key advantage of this method over other approaches is its ability to simultaneously measure the orientation and pattern speed of the bar, among other parameters, from individual simulation snapshots.
    As shown in Fig. \ref{fig:B1_parameters}b, the bar pattern speed remains nearly constant, decreasing only mildly from $30$ to $25 \, \text{km}\,\text{s}^{-1}\,\text{kpc}^{-1}$ in 561\,Myr.
    Once these values are obtained, we employ \texttt{AGAMA} \citep{Vasiliev2019} to construct the gravitational potential for each snapshot.
    We adopt the default parameters presented in \citet{Vasiliev2019} for the density and potential expansions, {and assume bisymmetry when computing the potential. This is motivated by the fact that the bar- and spiral-driven $m=2$ modes dominate over higher-order and odd modes, making this a good approximation for the purpose of constructing the model potential}.
    
    Note that the combination of these tools --namely \citet{Dehnen2022} method and \texttt{AGAMA} \citep{Vasiliev2019}-- allows us to compute the gravitational potential independently at each snapshot.
    {Each snapshot is treated as a time-independent system; however, by repeating the procedure for successive snapshots, we capture the temporal evolution of the galaxy through the corresponding changes in the potential. This effectively introduces time dependence into the study in the form of a sequence of quasi-static models, providing a physically grounded description of the simulated galaxy's evolution.} 
    
    After extracting these initial parameters, we then proceed with the computation of the equilibrium points and the subsequent construction of the invariant manifolds \citep{MRG2006, MRG2007}. 
    As the modelling of the gravitational potential in this work is two-dimensional, the subsequent analysis focuses on particles whose motion is predominantly planar.
    We therefore select those satisfying $\vert z\vert<400$ pc and  $\vert v_z\vert<20\, \text{km}\,\text{s}^{-1} \, \text{kpc}^{-1}$ in our results.
    {This selection is applied a posteriori and does not affect the construction of the potential or the computation of equilibrium points and manifolds. 
    Instead, it ensures consistency with the adopted planar approximation, as particles with large vertical excursions would probe regions where the potential is not accurately modelled and could therefore introduce systematic biases. 
    Thus, this selection is primarily implemented to properly characterise the planar dynamical behaviour of the galactic disc.}
    This selection retains roughly 55--65\% of the total disc population.

\section{Identification of spiral arm signatures}\label{sec:kinematic}

    \begin{figure}[!hb]
   \centering
   \includegraphics[width=\hsize]{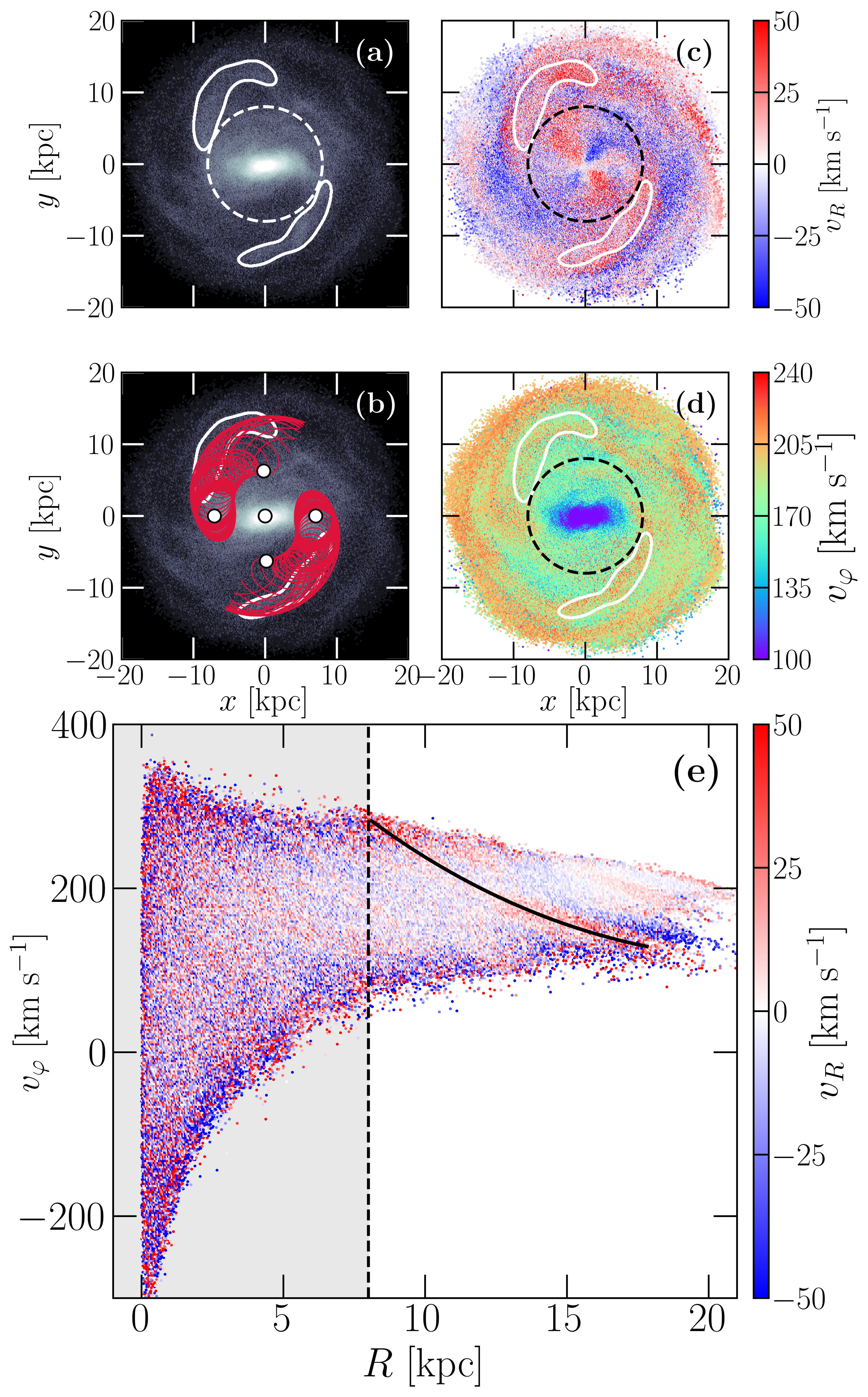}
        \caption{
        Face-on view and kinematic properties of the disc at $t=1.141$ Gyr after the initial conditions.
        \textit{(a)}: Face-on surface density map, with the spiral arm overdensities highlighted by white solid contours. 
        \textit{(b)}: Same as (a), with the exterior unstable manifolds overplotted in red.
        \textit{(c)}: Same as (a), colour-coded by radial velocity, $v_R$.
        \textit{(d)}: Same as (a), colour-coded by tangential velocity, $v_\varphi$.
        \textit{(e)}: Rotation curve, $v_\varphi$ as a function of radius, colour-coded by $v_R$. The solid black line indicates the ridge associated to the signature of the spiral arms. 
        The radial extent of the bar is indicated by dashed circles in the face-on panels, and by the grey shaded region together with the vertical dashed line in panel (e){, corresponding to $R_1 = 8.1\,\mathrm{kpc}$ at this time}.}
         \label{fig:vel_analytics}
   \end{figure}
    Across the selected snapshots of the simulation, two prominent, bisymmetric, grand-design spirals are clearly identifiable in the face-on view of the disc (Fig. \ref{fig:vel_analytics}a-d, white contours).
    {These contours trace the spatial extent of particles associated with the spiral arms, identified through a Difference-of-Gaussians (DoG) filtering of the stellar surface density.
    The DoG field is constructed from a two-dimensional histogram of the stellar distribution} by subtracting two Gaussian-smoothed maps with smoothing scales $\sigma_1 = 2$ and $\sigma_2 = 20$ histogram bins (corresponding to $\sim0.3$ and $\sim3.4$ kpc for the adopted map resolution), {and normalising by the large-scale component.
    Spiral regions are then defined as those above the $80^\text{th}$ percentile of the resulting field. The white contours are obtained from the spatial distribution of particles selected within these regions (and outside the bar region bounded by $R_1$).}
    This procedure enhances large-scale spiral features while suppressing small-scale fluctuations.
    
    Figure~\ref{fig:vel_analytics}b further shows that the exterior unstable branches of the invariant manifolds (red curves) associated with {the Lyapunov orbits around} the equilibrium points $L_1$ and $L_2$ (white dots {at the bar ends}) are geometrically consistent with the observed location of the spiral arms.
    For this snapshot (at $t=1.141$~Gyr after the initial conditions), the Jacobi energy of the saddle equilibrium points is $E_{L_{1,2}}=-1.65 \times 10^5\,(\text{km}\,\text{s}^{-1})^2$ and the energy of the manifold is $E_\text{man}=-1.57 \times 10^5\,(\text{km}\,\text{s}^{-1})^2$.
    {In addition to $L_1$ and $L_2$, the other three equilibrium points are also visible in Fig.~\ref{fig:vel_analytics}b, aligned with the bar’s semi-minor axis. 
    As expected, $L_3$ corresponds to a local minimum of the effective potential and is linearly stable, whereas $L_4$ and $L_5$ are located at local maxima and are also stable throughout the analysed time interval.  
    Therefore, in our model the dominant contribution to the spirals is directly provided by the manifolds associated with the Lyapunov orbits around $L_1$ and $L_2$.}
    
    Panels (c) and (d) of Fig.~\ref{fig:vel_analytics} present face-on views of the galaxy colour-coded by the radial and tangential velocities in the inertial frame, respectively. 
    In addition to the quadrupolar pattern in radial velocity within the bar region, the most striking feature in panel (c) is the presence of outward radial motion along the spiral arms (prominent red bands coincident with the white contours).   
    This radial velocity pattern has also been reported in previous works such as \citet{Eilers2020, Bernet2025}, and is consistent with the predictions of manifold-driven scenario \citep{Athanassoula2009a, Athanassoula2012}, while opposing to the expectations in classical density wave scenario \citep{LinShu1964, LinShu1966, LinShu1969, Kalnajs1973}.
    {In the framework of classical density wave theory, a steady spiral pattern in the absence of a bar is expected to display negative radial velocities inside corotation and positive radial velocities outside corotation \citep[e.g.][]{LinShu1969, Antoja2016, Bernet2025}.
    This behaviour is not observed in our case, where outward radial motions are found along the spiral arms over their full extent.
    Furthermore, following the azimuthal variation of radial velocities at fixed radius in Fig. \ref{fig:vel_analytics}c reveals that both the spiral and bar regions exhibit a similar red–blue–red–blue sequence, corresponding to a quadrupolar pattern, as also reported in previous works \citep[e.g.][]{Bovy2019, Ghosh2025}.
    The presence of this common quadrupolar signature in both components points to a shared dynamical origin. 
    In particular, this behaviour is naturally explained in the context of manifold theory, where spiral arms are generated by the bar and therefore inherit its kinematic imprint.}
    
    Panel (d) in Fig. \ref{fig:vel_analytics} shows that the spiral arms also exhibit enhanced tangential velocities compared to neighbouring disc regions at the same galactocentric radius.
    These kinematic signatures are combined in Fig.~\ref{fig:vel_analytics}e, where the $R$-$v_\varphi$ distribution is shown for all disc particles, colour-coded by radial velocity. Consistent with panels (c) and (d), a dominant red sequence in $v_R$ extends from the bar edge (vertical dashed line) out to $R \approx 20$~kpc, tracing the location and signature of the spiral arms (solid black line, Fig. \ref{fig:vel_analytics}e).
    Similar prominent $v_R>0$ ridges in the $R-v_\varphi$ plane also have been reported in barred galaxy models \citep[e.g.][]{Hunt2019, Fragkoudi2019, Khanna2019, Antoja2022}.

\section{Definition of different kinematic populations}\label{sec:populations}

\subsection{Jacobi energy of the manifolds, $E_\text{man}$}\label{sec:manifold_E}

We identify distinct kinematic populations using a criterion that is closely related to, but conceptually reformulated from, that introduced by \citet{KaufmannContopoulos1996}.
In their work, chaotic orbits were shown to support the inner parts of spiral arms provided that their Jacobi energy lies between the values at the Lagrange points $L_1$–$L_2$ and $L_4$–$L_5$.
{These orbits were already associated with the so-called ‘hot’ population identified in earlier self-consistent $N$-body simulations \citep{SparkeSellwood1987, PfenningerFriedli1991}, thereby placing them within a broader dynamical framework that is consistent with the present model.}

This energy range arises naturally  {in the presence of the rotating bar, which introduces a non-axisymmetric perturbation that separates the Jacobi energies of the equilibrium points $L_1$–$L_2$ and $L_4$–$L_5$.
No such separation occurs in the purely axisymmetric case \citep{Contopoulos2002}.
Within this energy interval, the topology of the zero-velocity curves allows for orbital configurations that can support spiral structures (see Appendix~\ref{sec:forbidden} for a detailed description).}

Building on this picture, we reinterpret and slightly reformulate the argument within the framework of invariant manifolds. 
{In practice, invariant manifolds are computed at the same Jacobi energy as their associated Lyapunov periodic orbits \citep{MRG2006}, so that each manifold is defined at a specific value of $E_J$.}
In particular, for a Lyapunov orbit $n$ around $L_1$ or $L_2$, {the associated invariant manifold exists at a fixed Jacobi energy $E_{\text{man}}^{(n)}$}, with $E_{L_{1,2}}<E_{\text{man}}^{(n)}<0$ \citep{Skokos2002, MRG2006}.
A natural choice would then be to impose an upper limit equal to $E_{L_{4,5}}$ at each individual snapshot.
However, in a fully self-consistent simulation the Jacobi energy of the equilibrium points evolves over time, because both the pattern speed  (Fig.~\ref{fig:B1_parameters}b) and the mass distribution of the system change.

To account for this evolution in a simple and robust way, we adopt a single reference energy for all snapshots. 
Specifically, we define $E_\text{man}=E_{L_{1,2}}+\Delta E$, where $\Delta E\equiv \max\left\{E_{L_{4,5}}^{(i)}-E_{L_{1,2}}^{(i)}\right\}$ is the maximum separation between $E_{L_{4,5}}$ and $E_{L_{1,2}}$ measured across all snapshots $i$.
In this way, instead of using a time-dependent upper limit, we fix a global one based on the largest energy gap reached during the evolution.
{This choice has two main advantages.
First, it ensures that the same energy range is used consistently at all times, avoiding artificial variations due to snapshot-to-snapshot fluctuations. 
Second, it provides a conservative definition: the selected range enforces a coherent and uniform energy gap at all times, thereby capturing the same physically relevant energies expected to contribute to spiral structure.}

{By construction, this global upper limit can be higher than the instantaneous value of $E_{L_{4,5}}$ at a given time. As a result, we may include manifolds with $E_J > E_{L_{4,5}}$. 
These manifolds are still dynamically allowed, but their ability to support spiral structure is expected to be weaker, in agreement with previous studies \citep[e.g.][]{KaufmannContopoulos1996}.
Importantly, this extension does not contradict the original criterion: for Jacobi energies above $E_{L_{4,5}}$, no forbidden regions exist (see Appendix~\ref{sec:forbidden}), so including these manifolds remains dynamically consistent with the criterion of \citet{KaufmannContopoulos1996}.}
Rather than modifying the underlying physics, our approach provides a practical generalisation that naturally incorporates the secular evolution of the system in a fully self-consistent $N$-body simulation while preserving the original dynamical constraints.

    \begin{figure}[!t]
   \centering
   \includegraphics[width=\hsize]{B1_ene_peq.png}
        \caption{Temporal evolution of the Jacobi energies at the unstable equilibrium points.
        \textit{(a)}: Jacobi energy at the Lagrange points $L_1$–$L_2$ (red) and $L_4$–$L_5$ (blue) as a function of time. Shaded regions indicate the dispersion associated with the determination of the Jacobi energies.
        \textit{(b)}: Energy separation $\Delta E_J = E_{L4,5} - E_{L1,2}$ as a function of time. The grey band marks the corresponding dispersion. The maximum value of this separation over the analysed time interval defines the constant energy offset $\Delta E$ adopted in our manifold selection criterion.}

         \label{fig:B1_ene_peq}
   \end{figure}

\subsection{Kinematic populations and their imprints}
\begin{figure*}[!h]
   \centering
   \includegraphics[width=0.92\hsize]{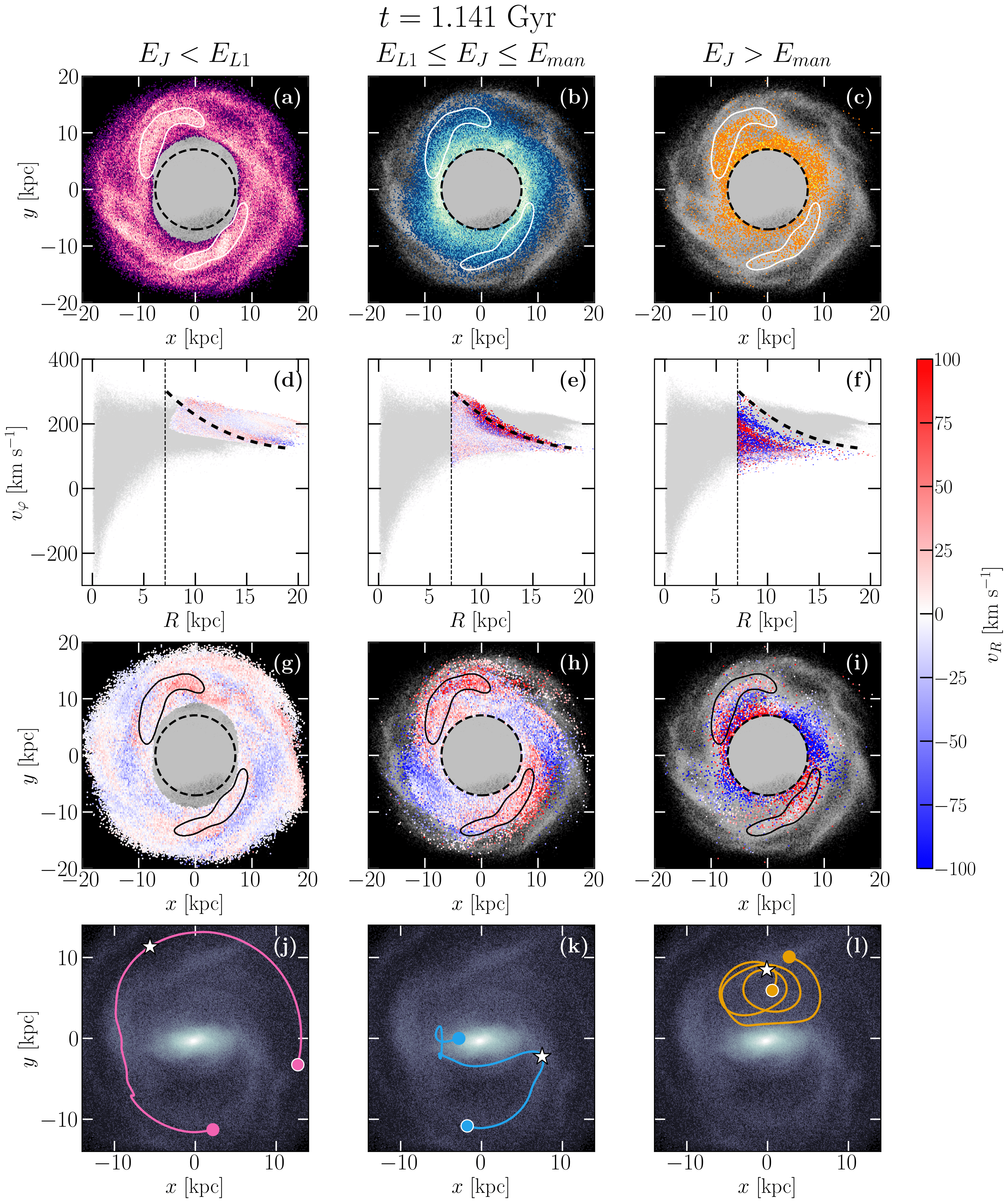}
        \caption{
        Dynamical diagnostics for the different kinematic populations at $t=1.141$ Gyr {after the beginning of the simulation}.
        \textit{Left column}: low-energy population.
        \textit{Middle column}: manifold-compatible population; 
        \textit{Right column}: high-energy population.
        \textit{First row}: Face-on particle distribution colour-coded by their surface density. The dashed circles indicate the extent of the bar.
        \textit{Second row}: Tangential rotation curve, $v_\varphi$ as a function of radius $R$, colour-coded by radial velocity $v_R$. The vertical dashed line marks the radial extent of the bar region. The grey background points correspond to the full distribution shown in Fig.~\ref{fig:vel_analytics}e and are included for reference.
        \textit{Third row}: Face-on particle distribution colour-coded by their radial velocity, $v_R$. The galaxy rotates counter-clockwise.
        The dashed circles indicate the extent of the bar.
        \textit{Fourth row}: Representative particle orbits overplotted on the face-on surface density map. A white star marks the position of the particle at $t=1.141$ Gyr. The start and end points of the orbit segment considered are indicated by solid circles, with the white-outlined one corresponding to the final time.
        }
         \label{fig:populations}
   \end{figure*}

Based on the criterion derived in Sec. \ref{sec:manifold_E}, all particles located outside the bar region
—identified by applying a radial mask $R>R_1$, where $R_1$ corresponds to the outer edge of the bar defined in \citet{Dehnen2022}, thus excluding bar particles from this analysis— can be classified according to their Jacobi energy, $E_J$ (defined in Eq. \eqref{eq:jacobi}), into three dynamically distinct populations:
(i) low-energy particles ($E_J<E_{L_{1,2}}$, Fig.~\ref{fig:populations}a),
(ii) manifold-compatible\footnote{As will be discussed in this section, this is a necessary but not sufficient condition to ensure that particles lie on manifold orbits; the details of the trapping mechanism will be presented in Paper II.} particles ($E_{L_{1,2}}\leq E_J \leq E_\text{man}$, Fig.~\ref{fig:populations}b), and
(iii) high-energy particles ($E_J>E_\text{man}$, Fig.~\ref{fig:populations}c).
Panels (d)–(f) of Fig.~\ref{fig:populations} present the same analysis shown in Fig.~\ref{fig:vel_analytics}e, now computed separately for each of these populations. Similarly, panels (g)–(i) reproduce the face-on view of the galaxy shown in Fig.~\ref{fig:vel_analytics}c, but restricted to each of the corresponding energy ranges. Panels (j)–(l) display representative orbits for each kinematic population.\\

First, {low-energy particles ($E_J<E_{L_{1,2}}$) can only access regions outside the zero-velocity curve associated with $E_{L_{1,2}}$, which outlines the innermost forbidden region for this energy range (see Appendix \ref{sec:forbidden} for more details).
As a result, they populate the vast majority of the disc, extending from this inner boundary} out to the disc edge, as shown in Fig. \ref{fig:populations}a.
Disc particles within this energy range do not have a sufficiently high Jacobi constant to cross the corotation radius, therefore they remain confined to the outer disc\footnote{In a similar way, particles in the bar region whose Jacobi energy is $E_J<E_{L_{1,2}}$ are also confined inside corotation.}, as also discussed in \citet{KaufmannContopoulos1996}.

As in Fig. \ref{fig:populations}a, the broad radial distribution of the low-energy population is also evident in Fig.~\ref{fig:populations}d, which additionally shows that their radial velocities are systematically small. 
This indicates that particles in this population follow nearly circular orbits. 
An illustrative example of a particle's orbits in the rotating frame is shown in Fig.~\ref{fig:populations}j.
Observe that, although the absolute values of the radial velocity are small, their spatial distribution is not random.
A coherent pattern is visible across the spiral arms: on the leading side\footnote{Recall the galaxy rotates counter-clockwise.} of the arms particles exhibit a systematic tendency toward negative radial velocities (Fig.~\ref{fig:populations}g), indicative of inward migration; whereas on the trailing side there is a preference for positive radial velocities, corresponding to outward migration.

This behaviour is also reflected in Fig.~\ref{fig:populations}d, where the migration signatures appear as secondary ridges located above the dominant ridge (marked by the dotted black line). 
These streaming motions can be naturally interpreted within the invariant manifold framework.
In this picture, low-energy particles do not generate the spiral pattern --like in the traditional density wave theory-- {but instead respond to the non-axisymmetric gravitational field generated by the spiral structure itself, arising from material channeled along the invariant manifolds}. 
Importantly, this leading–trailing asymmetry in the sign of the radial velocity has also been reported in studies such as \citet{SellwoodBinney2002, GrandKawata2016, SanchezMenguiano2018, Marques2025}, where radial migration was driven by transient spiral structure.
In the classical quasi-stationary density wave picture \citep{LinShu1964, LinShu1966, Weilen1974}, stars are expected to cross the spiral arms, with significant angular momentum changes occurring primarily near the corotation resonance.
By contrast, the streaming patterns observed here extend over a broader radial range and appear closely aligned with the phase-space geometry defined by the invariant manifolds, producing coherent inward and outward motions relative to the projection of the manifolds onto configuration space.\\

Second, manifold-compatible particles ($E_{L_{1,2}}\leq E_J \leq E_\text{man}$) originate near the bar region and extend to intermediate radii, tracing the inner regions of the disc, including the spiral arms (see Fig. \ref{fig:populations}b).
This panel shows that they do not populate the entire disc, although they are free to circulate throughout the bar region and into the outer disc within their energetically allowed domain (see Appendix \ref{sec:forbidden}).
In contrast to the low-energy population (Fig.~\ref{fig:populations}d), this group exhibits a clearly distinct dynamical behaviour. 
In Fig.~\ref{fig:populations}e, it produces a prominent red sequence corresponding to positive radial velocities, consistent with the outward radial motion previously identified in Fig.~\ref{fig:vel_analytics}c,e in the spiral arms.
This dominant ridge indicates large positive outward migration from the edge of the bar region to the outer disc.
This scenario is predicted by the transport of material within the invariant manifolds framework \citep[e.g.][]{MRG2006, MRG2007, Athanassoula2009a, Athanassoula2009b, Athanassoula2010, Athanassoula2012}.

However, energy compatibility with manifold orbits is only a necessary, but not sufficient, condition for particles to actually follow such trajectories. 
Many particles with energies within this range do not satisfy the additional dynamical conditions required to become trapped in manifold orbits, which explains why a clear spiral pattern is not traced by all particles in this energy interval. 
{In the present work, we do not attempt to identify which particles are dynamically guided by invariant manifolds at the level of individual orbits. Rather, we use this energy range as a kinematic selection that isolates particles potentially associated with manifold-driven structures.
A detailed characterisation of the dynamical conditions required for particles to effectively follow manifold-guided trajectories—such as phase-space accessibility, time-dependent transport, and trapping efficiency—will be addressed in Paper II.}

An example of a manifold-compatible orbit is presented in Fig.~\ref{fig:populations}h. 
Many of these trajectories originate in the bar region and subsequently travel along the invariant manifold, i.e. along the spiral arm. Such trajectories are referred to as \textit{transit orbits} \citep{MRG2006, MRG2007, Patsis2009, Patsis2010}.
In \citet{KaufmannContopoulos1996}, it was also found that orbits within the energetic bounds imposed by $L_1$–$L_2$ and $L_4$–$L_5$ could loop around the saddle points and eventually trace and support the spiral structure.

The adopted energy classification is therefore not merely descriptive, but dynamically meaningful. 
By separating particles according to their Jacobi energy relative to the values at $L_{1,2}$ and at the manifold, we isolate the subset of orbits that are dynamically compatible with the invariant manifolds. 
In particular, only particles within this manifold-compatible energy range can circulate around the saddle equilibrium points and subsequently follow the unstable manifolds that emanate from them. 
These are precisely the trajectories capable of producing genuine radial transport from the bar region toward the outer disc.
As shown in Fig. \ref{fig:populations}b, the surface density of this population is enhanced in the inner regions of the disc, where they connect to the bar.
The fact that the global density distribution of this population does not by itself reproduce the spiral pattern reflects the fact that the adopted energy range identifies a necessary dynamical condition for manifold-guided motion, but not a sufficient one. 
In other words, while all manifold-guided particles must belong to this energy class, not all particles within this class are necessarily trapped by the manifolds at a given time.
Within this manifold-compatible population lies the dynamically relevant subset of particles that can become trapped and stream along the manifolds, potentially mediating the effective redistribution of mass and angular momentum that underlies the spiral structure. 
In this sense, the manifold-compatible population defines a physically motivated set of candidates for generating the spiral pattern, from which the manifold-guided trajectories are drawn. 
Consequently, these constitute the dynamical backbone of the spiral structure in a barred galaxy.
The efficiency with which these particles are trapped by the invariant manifolds is expected to directly affect the prominence of the spiral arms: more efficient trapping would lead to a larger fraction of manifold-guided particles streaming along the arms, thereby increasing the density contrast produced by this population.
In the forthcoming Paper II, we will explicitly quantify the fraction of particles that are truly trapped by the invariant manifolds and follow transit orbits. 
This will allow for a direct comparison between the number of manifold-guided particles and the temporal evolution of the spiral arm strength, thereby providing a quantitative test of the manifold-driven scenario.\\
   
Finally, high-energy particles ($E_J>E_\text{man}$) appear scattered around the forbidden regions (Fig. \ref{fig:populations}c).
They exhibit large radial velocities, both positive and negative (see panels f and h in Fig. \ref{fig:populations}), as they correspond to banana orbits around the equilibrium points $L_4$ and $L_5$. 
Fig. \ref{fig:populations}l presents an example of these orbits.
As discussed in Appendix \ref{sec:forbidden}, there are no forbidden regions for Jacobi energies higher than $E_{L_{4,5}}$, hence particles in this regime are not subject to global spatial restrictions from an energetic standpoint, and can in principle explore the entire disc.
Their large radial motions are a natural consequence of this unrestricted accessibility.
The kinematic imprint of this population shown in Fig. \ref{fig:populations}f. 
When compared to  Fig.\ref{fig:vel_analytics}e, Fig. \ref{fig:populations}i indicates that there is little to none contribution of this population to the spiral configuration, at least directly.
In fact, this high-energy population brings about the ridges below to the dominant ridge associated with the bisymmetric spiral arms in the $R$-$v_\varphi$ plane.
It should be noted that the colours indicating $v_R$ in Fig. \ref{fig:populations}f,i are not weighted by number density, so features in those panels may appear strong even if there are very few stars, which is the case for the high-energy particles.\\

\begin{figure}[!b]
   \centering
   \includegraphics[width=\hsize]{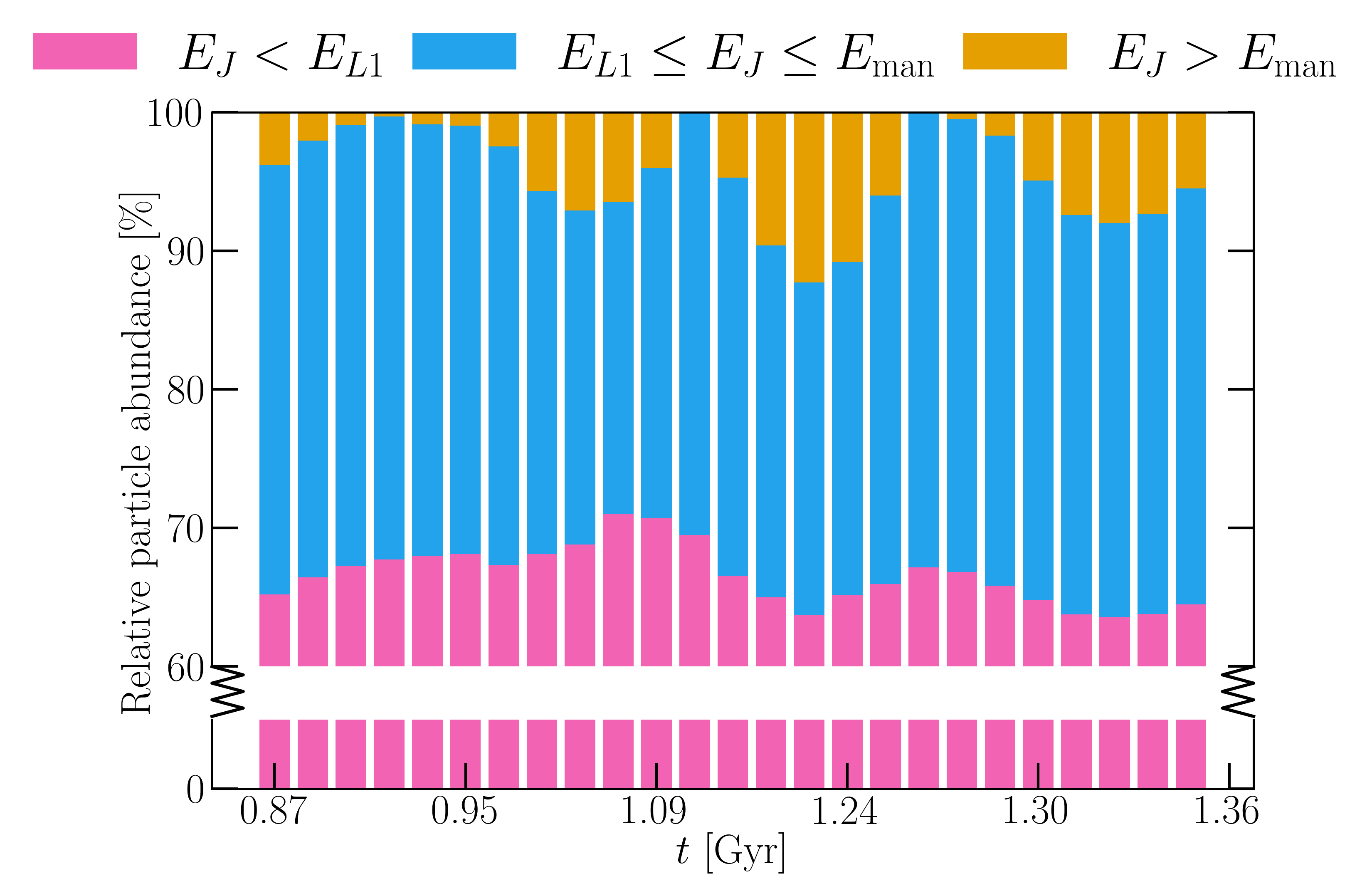}
        \caption{Temporal evolution of the relative contribution of each dynamical population, colour-coded as follows: low-energy (pink), manifold-compatible (blue), high-energy (orange).}

         \label{fig:B1_histo}
\end{figure}

The relative abundances of each of the three identified populations across all the simulation snapshots analysed are shown in Fig.~\ref{fig:B1_histo}. 
Low-energy particles (pink histogram) consistently account for approximately $\sim 60-70\%$ of the disc at all times, showing small variations in response to the strengthening or weakening of the spiral arms (Fig. \ref{fig:B1_parameters}a).
This indicates that they constitute the dominant and dynamically stable bulk of the disc. 
The high-energy population (orange histogram) never exceeds $\sim 10\%$, and in some cases falls to only a few percent, rendering its contribution marginal. 
The remaining $\sim 30$–$40\%$ of particles are manifold-compatible (blue histogram), which strongly reflect the variations in spiral arm amplitude.
This fraction of these particles is similar to the findings presented in \citet{Pfenninger1984b, PfenningerFriedli1991, KaufmannContopoulos1996}.

\section{Bar-driven origin of kinematic ridges}\label{sec:discussion}

{The clear association between kinematic populations and distinct ridge features in the $R$–$v_\varphi$ plane presented in Sec. \ref{sec:kinematic} strongly supports an interpretation in terms of orbital transport governed by invariant manifolds. 
Within this framework, the ridges are not merely passive imprints of phase mixing or transient perturbations; instead, they emerge as dynamical signatures of coherent orbital structures arising from energy differences between the Lagrangian points induced by the non-axisymmetric perturbation of the bar.}

{Each kinematic population can thus be understood as generating a distinct set of ridges. 
The manifold-compatible population gives rise to the prominent ridge associated with positive radial velocities, extending from the end of the bar to the outer disc. 
This intermediate energy range separates high- and low-energy populations, and the ridge structure reflects this distinction accordingly. 
Ridges below the dominant one are associated with banana orbits around $L_4$ and $L_5$, while those above it originate from mild deviations from circular motion within the low-energy population, driven by the self-gravity of the particles flowing along the spiral arms.
Altogether, this population-dependent stratification of ridges provides a coherent dynamical picture in which invariant manifolds act as the underlying phase-space skeleton organising both radial migration and the emergence of ridge-like features.}\\

{Although kinematic ridges were first identified observationally in the $R$–$v_\varphi$ plane of the Milky Way disc using high-precision astrometric data from the \textit{Gaia} mission \citep{Antoja2018}, similar phase-space structures have also been reported in numerous numerical simulations of barred galaxies \citep[e.g.][]{Antoja2018, Hunt2018b, Hunt2019, Martinez-Medina+2019, Fragkoudi2019, Monari2019, Khanna2019, Antoja2022, Bernet2022}.
This indicates that ridges are not exclusive to observational data of the Milky Way, but instead constitute a more general dynamical feature of barred systems, not necessarily restricted to Milky-Way-like galaxies or to purely observational contexts.}

{Regarding the dynamical role of the spiral structure, the present simulation exhibits two distinct regimes over the analysed time interval, as shown in Fig. \ref{fig:B1_parameters}a. 
In an initial phase, the amplitude of the spiral arms (as measured by the $A_2$ Fourier component) is comparable to that of the bar, indicating that the spirals constitute a dynamically significant perturbation. In a later phase, however, the spiral amplitude becomes subdominant relative to the bar, and in some snapshots the spiral structure nearly vanishes ($A_2^{\mathrm{spiral}} \lesssim 0.1$). 
Despite these differences, no significant changes are observed in the properties of the kinematic ridges between the two regimes. 
The ridge structures persist and are consistently identified both globally and within individual kinematic populations.}

{This result indicates that dynamically strong spiral arms are not a necessary condition for the formation of ridges in the $R$–$v_\varphi$ plane. 
This is consistent with previous studies showing that a rotating bar alone can generate similar ridge-like features \citep[e.g.][]{Antoja2018, Hunt2019, Monari2019, Bernet2022}. 
Our findings further support the scenario in which the bar is the primary driver of the observed phase-space structures, while spiral arms may modulate their properties yet are not required to for their formation.}

\section{Caveats and limitations}\label{sec:caveats}
{The results presented in this work are based on the analysis of a single $N$-body simulation snapshot sequence, and therefore their direct applicability is formally restricted to
systems sharing similar dynamical properties, as outlined in Sec. \ref{sec:methodology}.
Nevertheless, the theoretical framework underlying our interpretation—namely, the role of phase-space structures induced by non-axisymmetric perturbations—suggests that the main mechanisms identified here are expected to operate more generally in barred galaxies and barred–spiral systems. 
In this sense, while a broader exploration across different models would be required to fully assess the generality of our findings, the processes described are not specific to a unique realisation but are rooted in well-established dynamical principles.}\\

The main limitation of this study, as described in Sec.~\ref{sec:methodology}, lies in the planar nature of our analysis. 
The vertical cuts applied to the particle selection, namely $|z|<400\,\mathrm{pc}$ and $|v_z|<20\,\mathrm{km\,s^{-1}}$, effectively restrict the motion to be predominantly planar. 
We consider this approximation to be appropriate for the purposes of the present work, since the simulated galaxy is isolated and does not interact with any external perturber, which would otherwise introduce an additional layer of dynamical complexity.
{Moreover, previous studies of the three-dimensional dynamics in the vicinity of the Lagrange points have shown that the main orbital structures associated with invariant manifolds are largely preserved in the planar approximation, at least within the dynamical regimes considered here \citep{OllePfenniger1998, KatsanikasPatsis2022, HarsoulaKatsanikas2025}.}
Nevertheless, we acknowledge that a full characterisation including the vertical component would be required to provide a more realistic and complete picture of the structures associated with invariant manifolds.

A secondary limitation of our analysis concerns the modelling of the gravitational potential.
In this work, we analyse a sequence of snapshots from an $N$-body simulation that exhibit a prominent grand-design spiral structure, as shown in Fig.~\ref{fig:B1_parameters}a. 
Accordingly, the gravitational potential has been extracted using \texttt{AGAMA} \citep{Vasiliev2019} under the assumption of bisymmetry. 
{While this approximation neglects higher-order asymmetries in the density distribution, which are visible in the snapshots, we expect the resulting discrepancies in the potential to be of second order and not to significantly affect the main results of our analysis.}

\section{Conclusions}\label{sec:conclusions}

In this work, we have investigated the dynamical origin and kinematic signatures of spiral structure in an $N$-body simulation of an isolated barred galaxy. By combining the characterisation method of \citet{Dehnen2022} with snapshot-specific gravitational potentials constructed using \texttt{AGAMA} \citep{Vasiliev2019}, {we compute invariant manifolds in a sequence of quasi-static models that adapt self-consistently to the instantaneous state of the system}.
The main conclusions to be drawn from this study are summarised as follows:

\begin{itemize}
    \item[1.] The Jacobi constant $E_J$ provides a robust indicator in $N$-body simulations of the dynamical nature of disc particles {relative to the Jacobi energies that define the families of Lyapunov periodic orbits around the saddle equilibrium points $L_1-L_2$, whose associated invariant manifolds are explored in this work.} 
    We identify three dynamically distinct populations: 
    (i) low-energy particles, which dominate the disc ($\sim 60\%$; $E_J<E_{L_{1,2}}$) and follow nearly circular orbits; 
    (ii) high-energy particles, a minor component ($\lesssim 10\%$; $E_J>E_\text{man}$) associated with banana orbits around $L_4$ and $L_5$; and 
    (iii) manifold-compatible particles ($\sim 30$--$40\%$; $E_{L_{1,2}}\leq E_J\leq E_\text{man}$), which originate near the bar and can follow transit orbits along the spiral arms. 
    The latter population has been discussed in earlier studies \citep[e.g.][]{SparkeSellwood1987, Pfenninger1984b, PfenningerFriedli1991, KaufmannContopoulos1996}, and our findings are fully compatible with their conclusions. 
    In those works, however, the identification of this component was primarily based on indirect diagnostics and modelling strategies that did not allow for an explicit dynamical construction.
    In the present study, we adopt a fully reproducible framework grounded in invariant manifold theory and we apply it directly to an $N$-body simulation.
    This approach enables a systematic and transparent identification of the population, together with a detailed characterisation of its kinematic properties. 
    By recasting the phenomenon within the formalism of invariant manifolds, we provide a rigorous dynamical foundation that clarifies, unifies, and quantitatively substantiates the interpretations proposed in previous works. \\

    \item[2.] The theory of invariant manifolds provides a coherent dynamical interpretation of the ridge structure in the $R$–$v_\varphi$ plane of a barred galaxy with two grand-design spiral arms. 
    Within this framework, all ridges can be systematically classified according to their underlying energy populations, as suggested by \citet{Ramos2018, Khanna2019}.
    The dominant, long-lived ridge of positive radial motion extending from the bar to the outer disc is traced exclusively by the manifold-compatible population. 
    Only this component reproduces the characteristic kinematic signature of the bisymmetric spiral arms in the inertial frame, namely systematic outward motion ($v_R > 0$). 
    Secondary ridges below the main structure arise from the high-energy population, while those above it correspond to the low-energy population. \\

    \item[3.] Low-energy particles located near the projection of invariant manifolds onto configuration space crowd along the spiral structure, yet they do not actively shape its dynamics.
    This population does not intrinsically encode the dynamical properties of the spiral arms; instead, it represents the kinematic response to the gravitational influence of the manifold-compatible population that organises the large-scale spiral structure.
    A systematic streaming pattern is nevertheless evident: low-energy particles on the leading side of the arm migrate inward, whereas those in the trailing side migrate outward.
    This behaviour is induced by the combined influence of the invariant manifolds and the gravitational forcing exerted by the material flowing along them.
    {The small perturbations in radial velocity of these low-energy particles induced by the self-gravity of the spiral structure are broadly compatible with the ‘dimples’ of nearly-circular orbits in reported in \citet{KaufmannContopoulos1996} (see Fig. 19 there).
    In that work, the outer parts of the spiral arms were attributed to be sustained by the trapping of quasi-periodic orbits in the vicinity of stable periodic families beyond the $-4/1$ resonance of the bar.
    This raises the possibility that the perturbations identified here and the quasi-periodic trapping mechanism proposed in that work may coexist, contributing jointly to the structure of the outer spiral arms.
    However, in view of the kinematic imprints exhibited by this population (Fig.~\ref{fig:populations}g), the associated amplification of the density contrast appears to take place along the full extent of the spiral arm, rather than being confined to a specific radial or azimuthal segment.} 
    We then conclude that the primary contribution of this low-energy population is to amplify the density contrast of the spiral arms.
    Conversely, the manifold-compatible particles constitute the true dynamical backbone that governs the formation and evolution of the spiral structure, as they stream along the invariant manifolds and continuously replenish the arms.
    {This image is compatible with the results provided by \citet{KaufmannContopoulos1996}, where chaotic orbits, interpreted here as manifold-related trajectories, were invoked primarily to explain the inner part of the spiral arms.}\\
    
    \item[4.] We refer to this scenario as \textit{material density waves}.
    {We do not use the term ‘material’ to imply that the spiral arms follow the local rotation of the disc. 
    Instead, we use it to denote that the spiral structure is populated by particles that flow along it, rather than exclusively crossing it, since they are guided by the invariant manifolds that define the underlying dynamical backbone of the spiral arms.
    In this sense, the spiral structure combines a genuinely self-consistent dynamical component, associated with material guided by the manifolds flow, with a density-wave-like response of the surrounding disc.
    While this terminology may appear, at first sight, conceptually ambiguous—since it brings together notions that are often treated separately—we intentionally adopt it to highlight the hybrid nature of the mechanism we describe, in which organised particle transport along the invariant manifolds and the density-wave-like response are intrinsically linked.}

\end{itemize}

The systematic characterisation of all the ridge components of the $R$–$v_\varphi$ plane in terms of invariant manifolds is, to our knowledge, novel.
It demonstrates that only the manifold-compatible particles encode the intrinsic dynamical structure of the spiral arms in barred galaxies, while the remaining overdensities correspond to induced responses of the surrounding disc. 
In a forthcoming companion paper of this series (Paper II), we present a more detailed analysis of particle trapping by invariant manifolds and its relation to the time evolution of the spiral arms strength. 
This study explicitly quantifies the efficiency and duration of trapping episodes, and assesses how manifold-driven transport correlates with the strengthening and weakening of the spiral structure. 
We also investigate the differential impact of invariant manifolds on the various dynamical populations identified here, examining how each population contributes (or fails to contribute) to the growth, maintenance, and decay of the spiral arms.

\begin{acknowledgements}
{The authors would like to thank the anonymous referee for their exceptionally detailed and insightful comments, which helped to improve the clarity and quality of this work.}
The authors would also like to thank J.J. Masdemont and P. S\'anchez-Mart\'in for fruitful discussions about the manifold theory.
T.S.T. acknowledges that this work was partially supported by the FI-STEP predoctoral grant program of the Department of Research and Universities of the Generalitat de Catalunya, co-funded by the European Social Fund Plus (reference number: 2025 STEP 00196).
M.R.G. and T.S.T. acknowledge that this work was (partially) supported by the Spanish MICIN/AEI/10.13039/501100011033 and by "ERDF A way of making Europe" by the European Union through grants PID2021-122842OB-C21 and PID2024-157964OB-C21, the Institute of Cosmos Sciences University of Barcelona (ICCUB, Unidad de Excelencia Mar\'{\i}a de Maeztu) through grant CEX2024-001451-M and the project 2021-SGR-00679 GRC de l'Agència de Gestió d'Ajuts Universitaris i de Recerca (Generalitat de Catalunya).
S.R.F. acknowledges the finantial support by the Swedish National Space Agency Senior career grant with number 2025-00181, and the Spanish Ministry of Science and Innovation through the research grants: PID2021-123417OBI00, funded by MCIN/AEI/10.13039/501100011033/FEDER, EU; PCI2022-135023-2, funded by MCIN/AEI/10.13039/501100011033 and the EU “NextGenerationEU” / PRTR; and PID2024-157374OBI00, funded by MICIU/AEI/10.13039/501100011033/FEDER, EU. 
\end{acknowledgements}

\bibliographystyle{aa}
\bibliography{bibliography}

\begin{appendix}
   \section{Forbidden regions and ZVCs}\label{sec:forbidden} 
 
   As defined in \citet{MRG2006}, the gravitational potential of a barred galaxy in the rotating reference frame gives raise to \textit{forbidden regions}.
   These are not forbidden in an absolute sense for all particles; instead, they correspond to locations that a particle cannot access given its Jacobi energy, $E_J$.
   In this way, the distribution of particles across the galactic disc is strongly constrained by the forbidden regions defined with respect to their Jacobi constant.
   
\begin{figure}[!b]
       \centering
       \includegraphics[width=\hsize]{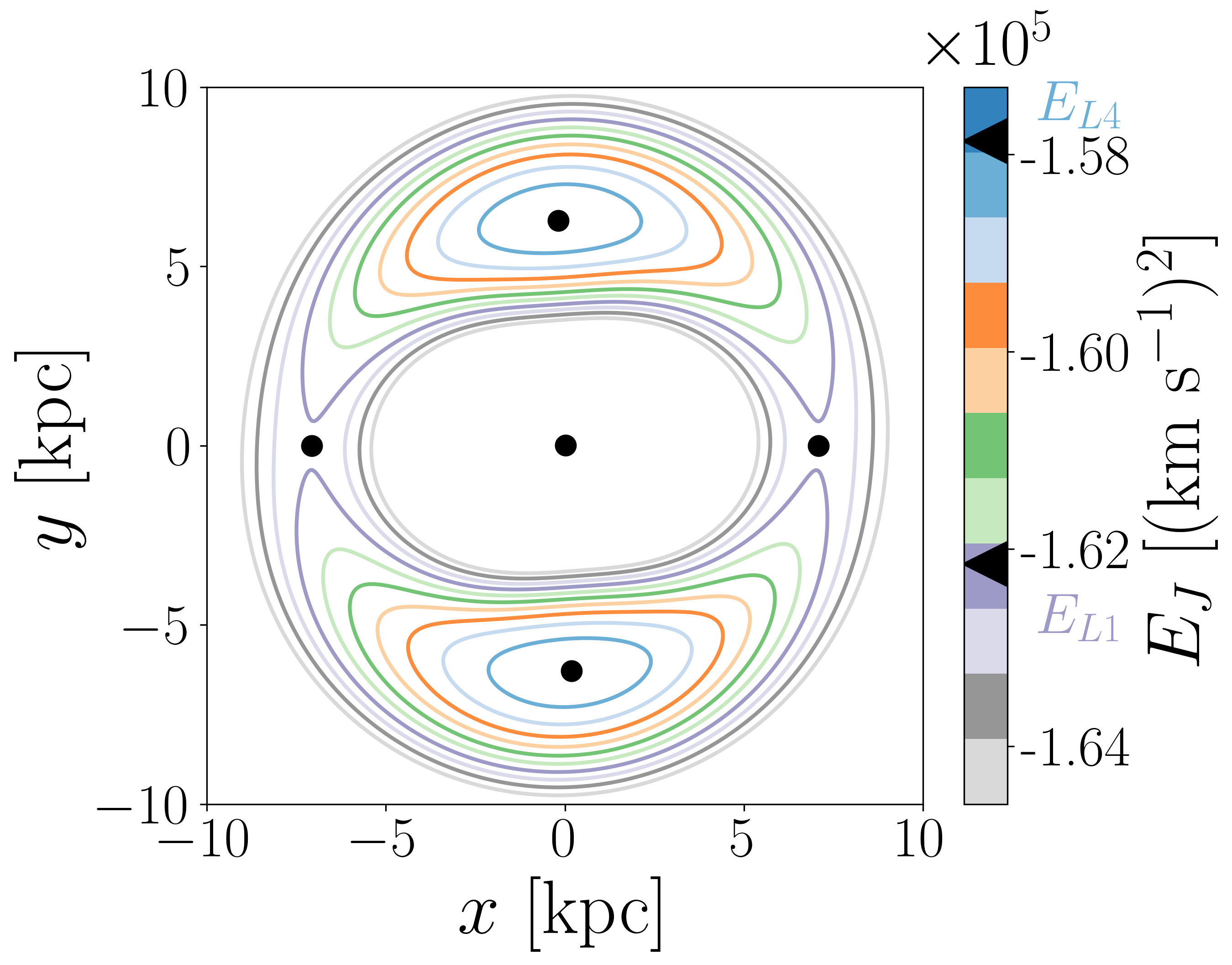}
            \caption{Zero velocity curves (ZVCs) corresponding to different Jacobi energies $E_J$, defined as the intersection of the surfaces $\Phi_\text{eff}=E_J$ with the plane $z=0$. 
            A discreet set of ZVCs is plotted from a selection of $E_J$, whose values are indicated in the colour bar, along the indication of $E_{L1}$ and $E_{L4}$ for reference.
            Black dots mark the location of Lagrange points.}
    
             \label{fig:B1_ZVC}
    \end{figure}

   The boundaries of these regions are determined by the zero-velocity surfaces, defined by the condition $\Phi_\text{eff}=E_J$.
   In practice, one often considers the zero-velocity curves (ZVCs), which correspond to the intersection of the zero-velocity surfaces with the galactic midplane.
   Importantly, both the size and morphology of the ZVCs depend on the value of $E_J$.
   For lower Jacobi energies (i.e. more negative $E_J$), the ZVCs enclose larger regions. 
   For $E_J=E_{L1}$, the two components of the ZVCs intersect at the Lagrange points $L_1$ and $L_2$.
   As $E_J$ increases, the curves progressively shrink, eventually collapsing onto the equilibrium points $L_4$--$L_5$.
   This behaviour is illustrated in Fig. \ref{fig:B1_ZVC}.
   In particular, for $E_J>E_{L4}\equiv E_J(L_4)$ no ZVCs exist, implying that the physical constraints imposed by the forbidden regions vanish.

   For the construction of the invariant manifolds, the relevant quantity is therefore the ZVCs associated with $E_{L1}\equiv E_J(L_1)$.
   From a dynamical perspective, $E_{L1}$ plays the role of a critical energy threshold that regulates orbital transport between the bar region and the outer disc. 
   For each $E_J<E_{L1}$, both components of the ZVCs remain closed respectively around the bar and at the outer disc, preventing orbits from traversing corotation. 
   Once $E_J$ exceeds $E_{L1}$, the ZVCs open at the saddle points, creating passages that allow material to flow between the inner and outer regions. In this sense, $E_{L1}$ acts as a dynamical gateway controlling the exchange of orbits across corotation.
   For $E_J>E_{L4}$, both the ZVCs and the forbidden regions disappear, rendering highly energetic particles spatially unconstrained.
   \begin{figure}[!h]
       \centering
       \includegraphics[width=0.53\hsize]{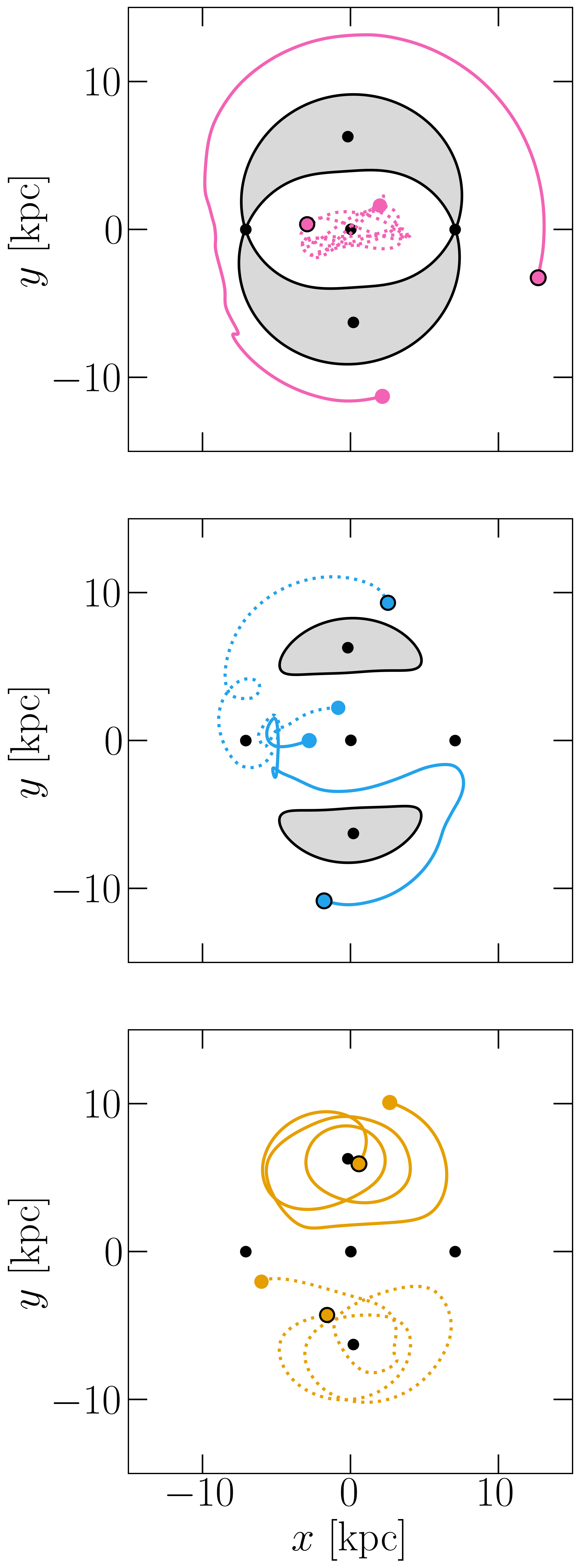}
            \caption{Forbidden regions (grey regions) and their corresponding zero-velocity curves (ZVCs, black contours) for different Jacobi constants: $E_J=E_{L1}$ (top), $E_{L1}<E_J<E_{L4}$ (middle), and $E_J=E_{L4}$ (bottom).
            Solid and dotted coloured curves correspond to two distinct orbits for the three different $E_J$ ranges defined in this work: low-energy (top), manifold-compatible (middle), high-energy (bottom); filled coloured circles mark their initial positions, while circles with black edges indicate their final positions in the analysed interval. 
            Black dots show the location of the equilibrium points.}
    
             \label{fig:B1_ZVC_orbi}
    \end{figure}
   This behaviour is illustrated in Fig. \ref{fig:B1_ZVC}. The top panel shows the configuration at the critical value $E_J=E_{L1}$, where the ZVCs are tangent at the saddle equilibrium points (this curve corresponds to the innermost region to which the low-energy population has access in Fig. \ref{fig:populations}a).
   Below this threshold, the allowed regions remain disconnected: orbits are still confined either to the bar region or to the outer disc, and large-scale radial transport is suppressed. 
   In the intermediate regime, $E_{L1}<E_J<E_{L4}$ (middle panel), the ZVCs open at the saddles, creating narrow channels that connect the inner and outer allowed regions. 
   These openings enable chaotic transport mediated by the invariant manifolds associated with the unstable periodic orbits around $L_1$ and $L_2$.
   As a result, stellar trajectories can now traverse corotation, moving between the bar and the outer disc along manifold-guided pathways.
   Finally, at $E_J=E_{L4}$ (bottom panel) the forbidden regions disappear, and the is no longer any constrain in the motion in configuration space. 
   Orbits become globally unbound within the rotating frame, exploring extended regions of the disc without energetic barriers. 
   This sequence clearly illustrates how the Jacobi constant governs the accessibility of phase-space regions and regulates the onset of large-scale orbital transport.
   {For further details, see Sec. 3.3.2 of \citet{BinneyTremaine} and Sec. 3.1.4 of \citet{Contopoulos2002}.}

\end{appendix}

\end{document}